\documentclass[aps,onecolumn,floats,prl]{revtex4}

\usepackage{amsmath}
\usepackage{amssymb}

\usepackage{graphicx}

\usepackage{natbib}

\usepackage{color} 

\usepackage[labelfont=bf,labelsep=period,justification=raggedright]{caption}

\makeatletter
\renewcommand{\@biblabel}[1]{\quad#1.}
\makeatother

\newcommand {\e}[1]{\mathrm{~#1}}       
\newcommand {\E}[1]{\cdot 10^{#1}}		

\def\urltilda{\kern -.15em\lower .7ex\hbox{\~{}}\kern .04em}
\def\urldot{\kern -.10em.\kern -.10em}
\def\urlhttp{http\kern -.10em\lower -.1ex\hbox{:}\kern -.12em\lower 0ex\hbox{/}\kern -.18em\lower 0ex\hbox{/}}

\begin{document}

\title{Natural images from the birthplace of the human eye}

\author{Ga\v{s}per Tka\v{c}ik$^{1,2,\ast}$}
\author{Patrick Garrigan$^{3}$}
\author{Charles Ratliff$^{4}$}
\author{Grega Mil\v{c}inski$^{5}$}
\author{Jennifer M. Klein$^{7}$}
\author{Lucia H. Seyfarth$^{6}$}
\author{Peter Sterling$^{6}$}
\author{David Brainard$^{7}$}
\author{Vijay Balasubramanian$^{1}$}

\affiliation{\mbox{
${}^1$ Department of Physics and Astronomy, University of Pennsylvania, Philadelphia, PA 19104, USA}
\\
\mbox{${}^2$ Institute of Science and Technology Austria, Am Campus 1, A-3400 Klosterneuburg, Austria}
\\
\mbox{${}^3$ Department of Psychology, Saint Joseph's University, Philadelphia, PA 19131, USA}
\\
\mbox{${}^4$ Department of Ophthalmology, Northwestern University, Chicago, IL 60611, USA}
\\
\mbox{${}^5$ Sinergise d.o.o, SI-1000 Ljubljana, Slovenia}
\\
\mbox{${}^6$ Department of Neuroscience, University of Pennsylvania School of Medicine, Philadelphia, PA 19104, USA}
\\
\mbox{${}^7$ Department of Psychology, University of Pennsylvania, Philadelphia, PA 19104, USA}
\\
$\ast$ E-mail: gtkacik@ist.ac.at
}

\begin{abstract}
Here we introduce a database of calibrated natural images publicly available through an easy-to-use web interface. Using a Nikon D70 digital SLR camera, we acquired about $5000$ six-megapixel images of Okavango Delta of Botswana, a tropical savanna habitat similar to where the human eye is thought to have evolved. Some sequences of images were captured unsystematically while following a baboon troop, while others were designed to vary a single parameter such as aperture, object distance, time of day or position on the horizon. Images are available in the raw RGB format and in grayscale. Images are also available in units relevant to the physiology of human cone photoreceptors, where pixel values represent the expected number of photoisomerizations per second for cones sensitive to long (L), medium (M) and short (S) wavelengths. This database is distributed under a Creative Commons Attribution-Noncommercial Unported license to facilitate research in computer vision, psychophysics of perception, and visual neuroscience. 


\end{abstract}

\maketitle

\section{Introduction}
High-resolution digital cameras are now ubiquitous and affordable, and are increasingly incorporated into portable computers, mobile phones and other handheld devices. This accessibility has led to an explosion of online image databases, accessible through photo-sharing websites such as Flickr and SmugMug, social networks such as Facebook, and many other internet sites. Disciplines such as neuroscience, computer science, engineering and psychology have profited from research into the statistical properties of natural image ensembles \cite{Simoncelli2001,Geisler2009}, and it might seem that the side benefit of public photo-sharing websites is an ample supply of image data for such research. Some research, however, requires carefully calibrated images that accurately represent the light that reaches the camera sensor. For example, to address questions about the early visual system, images should not deviate systematically from the patterns of light incident onto, and encoded by, the retina. Specific limitations of uncontrolled databases that hinder their use in vision research include {\bf (i)} compression by lossy algorithms, which distort image structure at fine scales; {\bf (ii)} photography with different cameras, which results in unpredictable quality and incomparable pixel values; {\bf (iii)} photography with different lenses or focal lengths, which may introduce unknown optical properties to the image incident on the sensor, and {\bf (iv)} photography for unspecified purposes, which may bias image content toward faces, man-made structures, panoramic landscapes, etc., while under-representing sky, ground, feces, or other more mundane content.

Due to these limitations, research involving natural images typically relies on a well-calibrated database \cite{Olmos:Kingdom:2004,VanHateren:1992,Parraga}. Examples where good image databases are essential include early visual processing in neural systems~\cite{Atick:1992p1283,Field:1987p1178,VanHateren:1992,Doi:Balcan:2007,Karklin:Lewicki:2009,Ratliff2010,Garrigan2010}, computer vision algorithms~\cite{DARPA,Forsyth:2002,Torralba:2003}, and image compression/reconstruction methods~\cite{jpeg,Simoncelli:2000}.  In each of these fields,  much of the research requires accurate characterization of the statistics of natural image ensembles.  Natural images exhibit characteristic luminance distributions~\cite{Richards:1982} and spatial correlations~\cite{Field:1987p1178,Burton:Moorhead:1987}.   These result from environmental regularities such as physical laws of projection and image formation, natural light sources, and the reflective properties of natural objects. Further, natural images exhibit higher-order regularities such as edges, shapes, contours and textures that are perceptually salient, but difficult to quantify \cite{Olshausen:1996,Bell:1997,Vorhees:1998,Riesenhuber:2000,Geisler:Perry:2001,Tkacik2010}. To characterize these regularities in image patches of increasing  spatial extent, the number of  image samples required to collect reliable joint pixel luminance probability distributions scales exponentially with the number of pixels in the patch, ultimately limiting the reliability of our estimates. Nevertheless, with sizable databases one might push the sampling limit to patches of up to $\sim 10\times 10$ pixels discretized into a few luminance levels, and even at that restricted size interesting results have emerged~\cite{Kersten:1987p102,Chandler:2007p16,Stephens:2008}. 

In this paper we describe a collection of calibrated natural images, which tries to address the shortcomings of uncontrolled image databases, while still providing substantial sampling power for image ensemble research. The database  is organized into themed sub-collections and is broadly annotated with keywords and tags.  The raw images are linear with incident light intensity in each of the camera's three color channels (Red, Green, and Blue). Further, because much research is concerned with human vision, we have translated each image into physiological units relevant to human cone photoreceptors. The raw image data, demosaiced RGB images, grayscale images representing luminance, and images in the physiologically relevant cone representation are available in the database.

\section{Results}
\subsection{Image calibration}
We checked {\bf (i)} that the Nikon D70 camera has a consistent resolution (in pixels per degree) in the vertical and horizontal directions; {\bf (ii)} that its sensor responses are linear across a large range of incident light intensities; and {\bf (iii)} that its shutter, aperture, and ISO settings behave regularly, so that it is possible to estimate incident photon flux given these settings.  We {\bf (iv)} measured the spectral response of the camera's R, G, and B sensors by taking images of a reflectance standard illuminated by a series of 31 narrowband light sources and verified that these measured responses allowed us to predict the R, G, and B responses to spectrally broadband light.  We also {\bf (v)} characterized the ``dark response'' of the camera, i.e. the sensor output with no light incident onto the CCD (charge-coupled device) chip.  Finally, we {\bf (vi)} measured the spatial modulation transfer function of each of the R, G, and B sensor planes for broadband incident light.  We examined two separate D70 cameras, and found that they yielded consistent results and could be used interchangeably, except for a single constant scaling factor in the CCD sensor response.  Taken together, the camera measurements allow us to transform raw camera RGB values into standardized RGB response values that are proportional to incident photon fluxes seen by the three camera sensor types.   Details on the camera characterization are provided in \emph{Materials and Methods: Camera response properties}.  

With these results in hand, we can further transform the standardized RGB data into physiologically relevant quantities (see details in \emph{Materials and Methods: Colorimetry}). The photon flux arriving at the human eye is filtered by passage through the ocular media before entering a cone photoreceptor aperture. There, with some probability, each photon may cause an isomerization. There are three classes of cones, the L, M, and S cones. Each contains a different photopigment, with the spectral sensitivities of the three photopigments peaking at approximately  $\lambda_S=421\e{nm}$, $\lambda_M=530\e{nm}$, and $\lambda_L=559\e{nm}$, respectively~\cite{Stockman:Sharpe:2000}.  We have transformed the RGB images into physiologically-relevant estimates of the isomerization rates that the incident light would produce in human L, M, and S cones.  For this transformation, we used the Stockman-Sharpe/CIE 2-degree (foveal) estimates of the cone fundamentals \cite{Stockman:Sharpe:2000, CIE2007}, together with methods outlined by Yin et al. \cite{Yin2006} to convert nominal cone coordinates to estimates of photopigment isomerization rates.  We have also transformed the RGB images into grayscale images representing estimates of the luminance of the incident light (in units of candelas per meter squared with respect to the CIE 2007 two-degree specification for photopic luminance spectral sensitivity).  An important contribution of this work is to report image data from a biologically relevant environment in units of cone photopigment isomerizations, thus characterizing the information available at the first stage of the human visual processing pathways.

\subsection{Image ensemble}

We have assembled a large database of natural images acquired with a calibrated camera (see Materials and Methods), and made it available to the general public through an easy-to-use web interface. The extensive dataset covers a single environment: a rich riverine / savanna habitat in the Okavango delta, Botswana, which is home to the full panoply of vertebrate and invertebrate species, such as lion, leopard, cheetah, elephant, warthog, antelope, zebra, giraffe, various bird species etc. This environment was chosen because it is thought to be similar to the environment where the human eye, and retina in particular, have evolved.  The database consists of $\sim 5000$ images, organized into about 100 folders (albums).  

\begin{figure}[!ht] 
   \centering
   \includegraphics[width=6in]{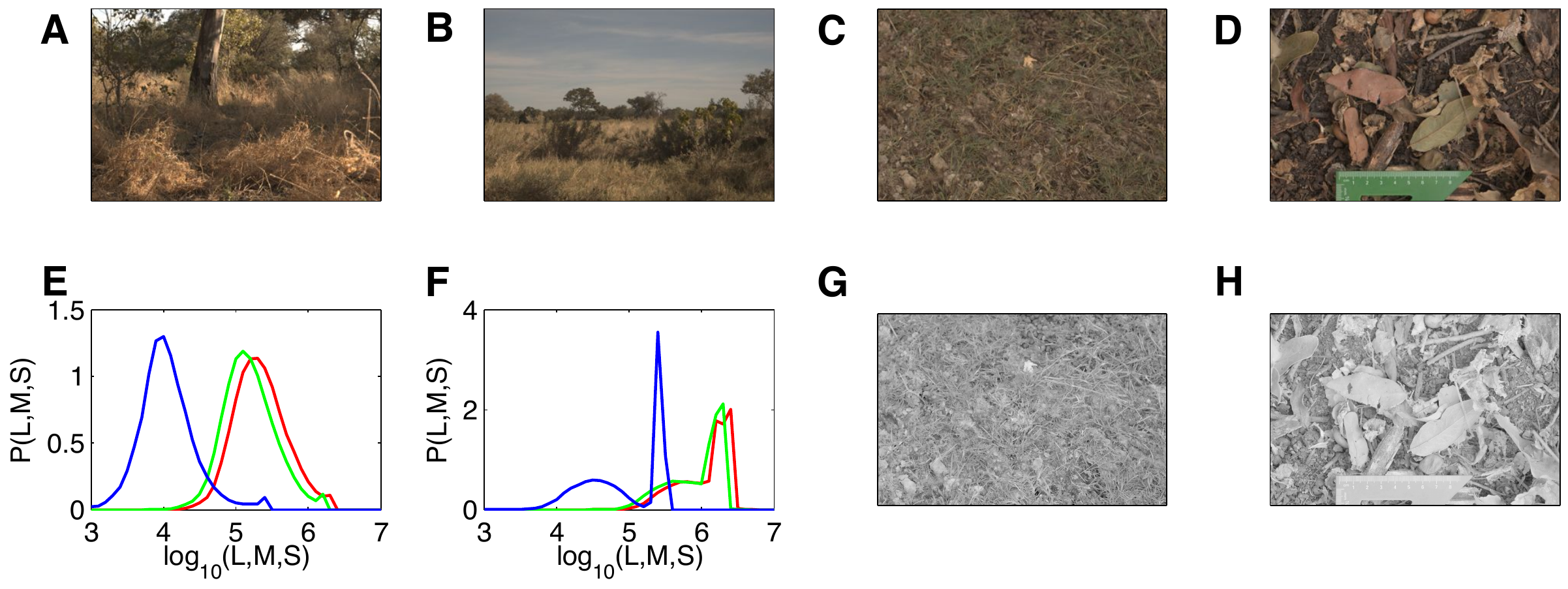}
   \caption{{\bf Example images from the Botswana dataset.} {\bf A-D)} Some natural scenes from various albums, including a tree, grass and bushes environment, the horizon with a large amount of sky, and closeups of the ground; the last image is from the image set containing a ruler than can be used to infer the absolute scale of objects. {\bf E-F)} The distributions of L (red), M (green) and S (blue) channel intensities across the image for images A) and B), respectively. The large sky coverage in B) causes a peak in the S channel at high values. The horizontal axis is log base 10 of pigment photoisomerization rates per second. {\bf G-H)} Grayscale images showing log luminance corresponding to the images in C) and D), respectively. } 
 \label{f-3}
\end{figure}

Tables \ref{t-1} and \ref{t-2} provide an album-by-album summary of the database content, along with content keywords for each album and tags that give additional information about how the images were gathered. Figure \ref{f-3} shows several examples of images from the database: baboon habitat, panoramic images that include the horizon, closeups of the ground, and closeup images that include the ruler for absolute scale determination.

\begin{figure}[!ht] 
   \centering
   \includegraphics[width=4in]{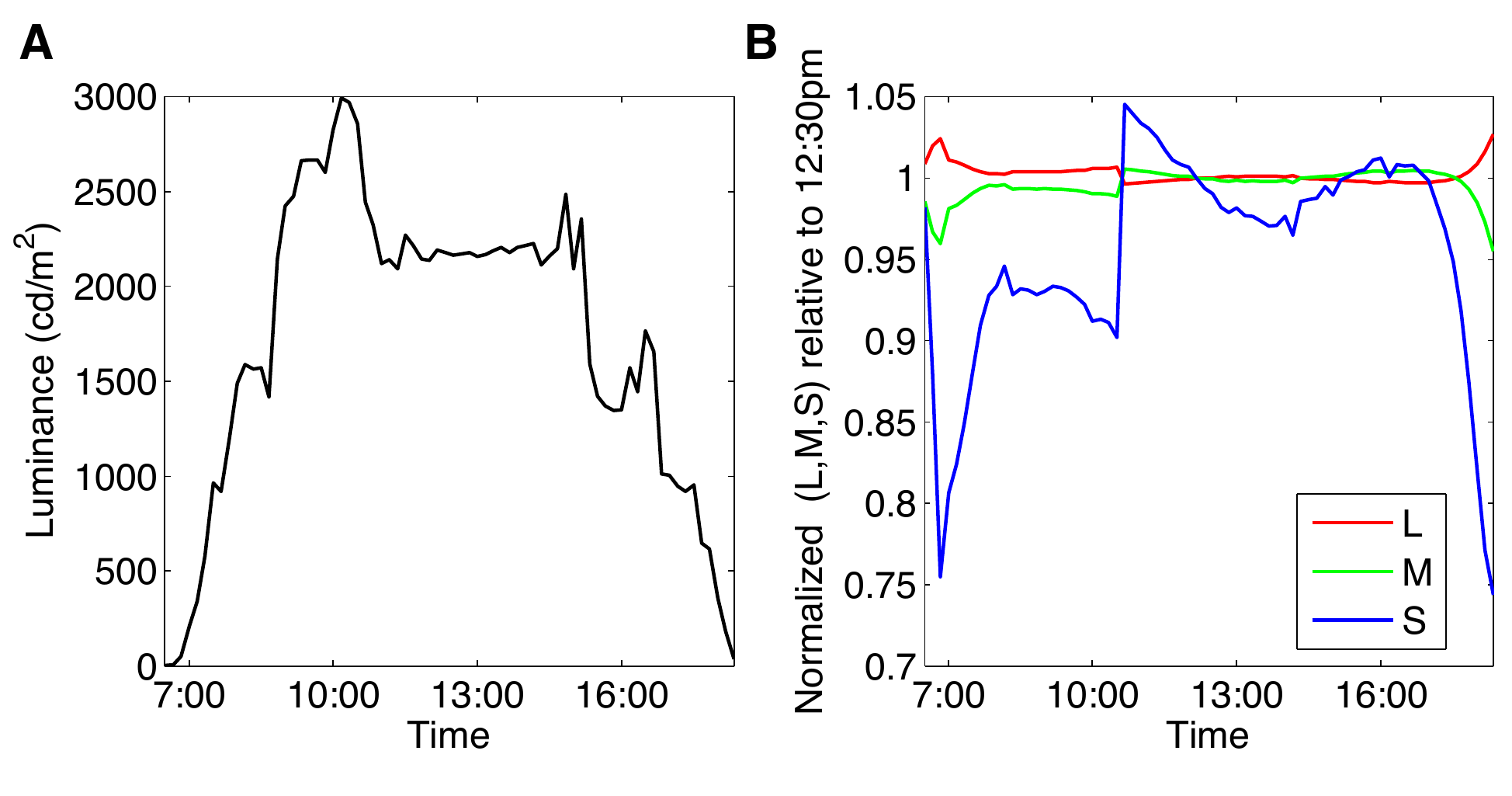}
   \caption{{\bf The color content and luminance of the sky.} {\bf A)} The  luminance in candelas per square meter shows the rise in the morning and decay in the evening, along with the fluctuations during the day. {\bf B)} The color content of the sky. To report relative changes in color content corrected for overall luminance variation, the L, M and S channels have been divided by their reference values at 12:30pm to bring the three separate curves together at the 12:30pm time point. In addition, all three curves were multiplied by the luminance at 12:30pm and then divided by the luminance at the time each measurement was taken.  At sunrise and sunset, the L (redder) channel is relatively more prominent.  } 
 \label{f-4}
\end{figure}

Figure \ref{f-4} shows a simple analysis of images from album {\sc cd32b}. Images of cloudless sky were taken every 10 minutes, from 6:30 until 18:30. We compute the average luminance as a function of time, and the LMS color composition of the light incident on the camera that was pointing up at the homogenous sky.

\begin{figure}[!ht] 
   \centering
   \includegraphics[width=3in]{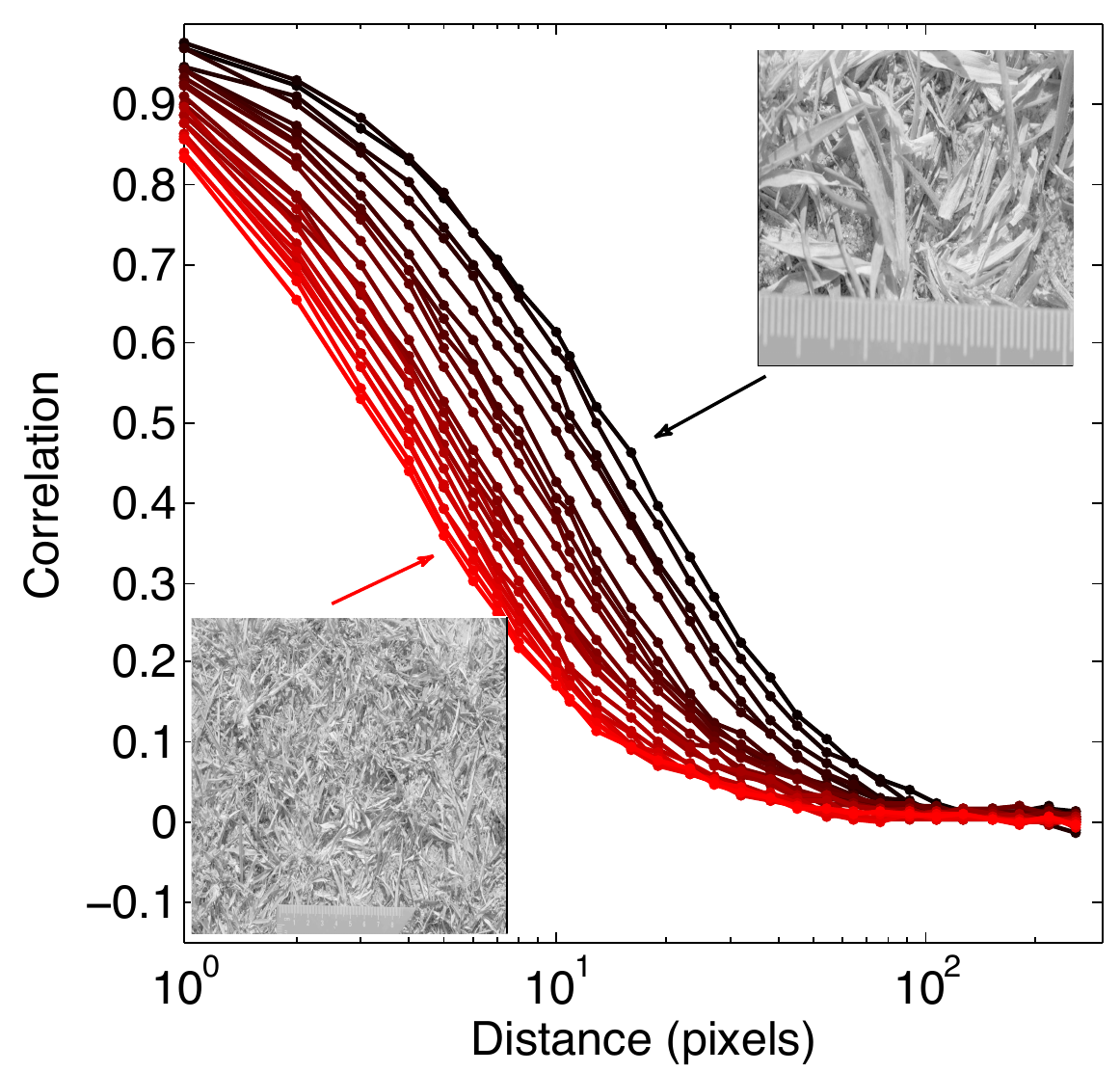}
   \caption{{\bf Pairwise correlations in natural scenes.} We analyzed 23 images of the same grass scrub scene, taken from different distances (black -- smallest distance,  red -- largest distance). For every image, we computed the pixel-to-pixel correlation function in the luminance channel, and normalized all correlation functions to be 1 at $R=0$ pixels. For largest distances, $R=256$ pixels, the correlations decay to zero. The decay is faster in images taken from afar (redder lines, the largest distance image shown as an inset in the lower left corner), than in images taken close up (darker lines, the smallest distance image shown as an inset in the upper right corner). All images contain a green ruler that facilitates the absolute scale determination; for this analysis, we exclude the lower quarter of the image so that the region containing the ruler is not included in the sampling.} 
 \label{f-5}
\end{figure}

Figure \ref{f-5} shows an analysis of 23 images from album {\sc cd04b}, where closeups of the grass scrub on the ground were taken from various distances. We computed the pairwise luminance correlation function as a function of pixel separation, and show how it varies systematically with distance to the ground. Since the grass scrub on our images has a preferred scale, the correlation function should decay faster the farther away from the scene the camera is, and this is indeed what we see. Scale invariance in natural scenes presumably emerges because of the distribution of object sizes and distances from which the objects are viewed \cite{lmh}. Our image collection can be used to study the scaling properties of the ensemble systematically.

The two examples provided above are intended to illustrate the strengths of our database: calibration and normalization into physiologically relevant units; organization into thematic (keywords) and methodological sequences (tags), that explore variations between the scenery content and variations induced by systematic changes in controllable parameters; and a thorough sampling that should suffice for the estimation of higher-order statistics or even accumulation of image patch probability distributions.

In assembling the database, we could have taken an alternative approach and  built a catalogue of as many objects from the environment as possible, photographed at a chosen ``standard'' camera position and controlled illumination, removed from the natural context and placed on a neutral background. While there are advantages to exhaustively pursuing this approach, we regarded it as focusing on natural \emph{objects} and not on natural \emph{scenes}. Nevertheless we provide a limited set of albums in the collection of Table \ref{t-2}, where closeups of fruit, grass, bark are provided; we reasoned that some of these closeups would be of use for studying properties of natural objects such as texture.

\subsection{Data access}
The image database is accessible at {\tt \urlhttp tofu\urldot physics\urldot upenn\urldot edu/\urltilda upennidb}, or through anonymous FTP at {\tt  ftp://anonymous@tofu\urldot physics\urldot upenn\urldot edu/fulldb}. A standard gallery program for viewing the images on the web makes browsing and downloading individual images or whole albums easy \cite{gallery}. Once images have been added to the `cart' and the user selects the download option, the database prompts the user to select the formats for download; the available formats are {\bf (i)} NEF (raw camera sensor output, Nikon proprietary format); {\bf (ii)} RGB Matlab matrix (demosaiced RGB values before dark response subtraction); {\bf (iii)} LMS Matlab matrix (physiological units of isomerizations per second in L, M, S cones); {\bf (iv)} LUM Matlab matrix (grayscale image in units of cd/m$^2$); and {\bf (v)} AUX meta-data Matlab structure (containing camera settings and basic image statistics). \emph{Materials and Methods: Image extraction and data formats} documents the detailed image processing pipeline. After image format selection has been made, the database prepares a download folder which contains the selected images in the requested formats and with the album (directory) structure maintained. The folder is available for download by (recursive) FTP, for instance by using an open-source {\sc wget} tool \cite{wget}; in this way large amounts of data, e.g., the whole database ($\sim 300\e{Gb}$ of image data for all available formats combined) can be reliably transferred. A list of requested images is provided with the image files, and can be used to retrieve the same selection of images from the database directly; this facilitates the reproducibility of analyses and uniquely defines each dataset. All images are distributed under Creative Commons Attribution-Noncommercial Unported license \cite{license}, which prohibits commercial use but allows free use and information sharing / remixing as long as authorship is recognized.  Images and processing software used to calibrate the cameras is available upon request.

\section{Materials and Methods}

{\bf Camera response properties.} Two D70 (Nikon, Inc., Tokyo, Japan) cameras are described. Each was equipped with an AF-S DX Zoom-Nikkor 18-70mm f/3.5-4.5G IF-ED lens. To protect the front surface of the lens, a 52 mm DMC  (Digital Multi Coated) UV skylight filter was used in all measurements. One camera, referred to in this document as the \emph{standard camera} (serial number {\tt 2000a9a7}), is characterized in detail. A subset of the measurements was made with a second D70 camera, referred to in this document as the \emph{auxiliary camera} (serial number {\tt 20004b72}). Its response properties matched those of the standard camera after all responses were multiplied by a single constant.

\emph{Geometric information.} The D70's sensor provides a raw resolution of $3040 \mathrm{(h)} \times 2014 \mathrm{(v)}$ pixels. The angular resolution of the camera was established by acquiring an image of a meter stick at a distance of 123 cm.  Both horizontal and vertical angular resolutions were 92 pixels per degree. This resolution is slightly lower than that of foveal cones, which sample the image at approximately 120 cones per degree~\cite{Packer}.

\emph{Image quantization.} The camera documentation indicates that the D70 has native 12-bit-per-pixel intensity resolution \cite{nikondoc1}, and that the NEF compressed image format supports this resolution. Nonetheless, the D70 appears not to write the raw 12-bit data to the NEF file. Rather, some quantization/compression algorithm is applied which converts the image data from $12$ bit to approximately $9.4$ bit per pixel resolution\cite{nikondoc2}. The raw data extraction program, {\sc dcraw}\footnote{We used version $v5.71$ of {\sc dcraw}; we note that the output image representation produced by this software is highly version dependent.} \cite{dcraw}, appears to correct for any pixel-wise nonlinearity introduced by this processing, but it cannot, of course, recover the full 12-bit resolution. The pixel values in the file extracted by {\sc dcraw} range between 0 and 16384 for red and blue, and between 0 and 16380 for the green channel, and these are the values we use for the analysis. We do not have a direct estimate of the actual precision of this representation.

\emph{Mosaic pattern and block averaging.} The D70 employs a mosaic photosensor array to provide RGB color images. That is, each pixel in a raw image corresponds either to an R, G, or B sensor, as shown in Fig \ref{f-2}. R, G, and B values can then be interpolated to each pixel location.  This process is known as demosaicing.  The images in the database were demosaiced by taking each $2$ by $2$ pixel block and averaging the sensors responses of each type with that block (one R sensor, 2 G sensors, and 1 B sensor).  This produced demosaiced images of $1519 \mathrm{(h)}\times 1007 \mathrm{(v)}$ pixels; these images were used in all measurements reported below.
\begin{figure}[]
\centering
\begin{tabular}{ccccccc}
\textcolor{blue}{B} & \textcolor{green}{G} & \textcolor{blue}{B} & \textcolor{green}{G} & \textcolor{blue}{B} & \textcolor{green}{G} & $\cdots$\\
\textcolor{green}{G} & \textcolor{red}{R} & \textcolor{green}{G} & \textcolor{red}{R} & \textcolor{green}{G} & \textcolor{red}{R} & $\cdots$ \\
\textcolor{blue}{B} & \textcolor{green}{G} & \textcolor{blue}{B} & \textcolor{green}{G} & \textcolor{blue}{B} & \textcolor{green}{G} & $\cdots$\\
\textcolor{green}{G} & \textcolor{red}{R} & \textcolor{green}{G} & \textcolor{red}{R} & \textcolor{green}{G} & \textcolor{red}{R} & $\cdots$ \\
$\cdots$& & & & & & 
\end{tabular}
\caption{A fragment of the mosaic pattern that tiles the CCD sensor. Each pixel is either red (R), green (G), or blue (B). The pixels are present in ratios 1:2:1 in the CCD array. The upper-left hand corner of the fragment matches the upper-left hand corner of the raw image data decoded by {\sc dcraw}.}
\label{f-2}
\end{figure}

\begin{figure}[!ht] 
   \centering
   \includegraphics[height=2in]{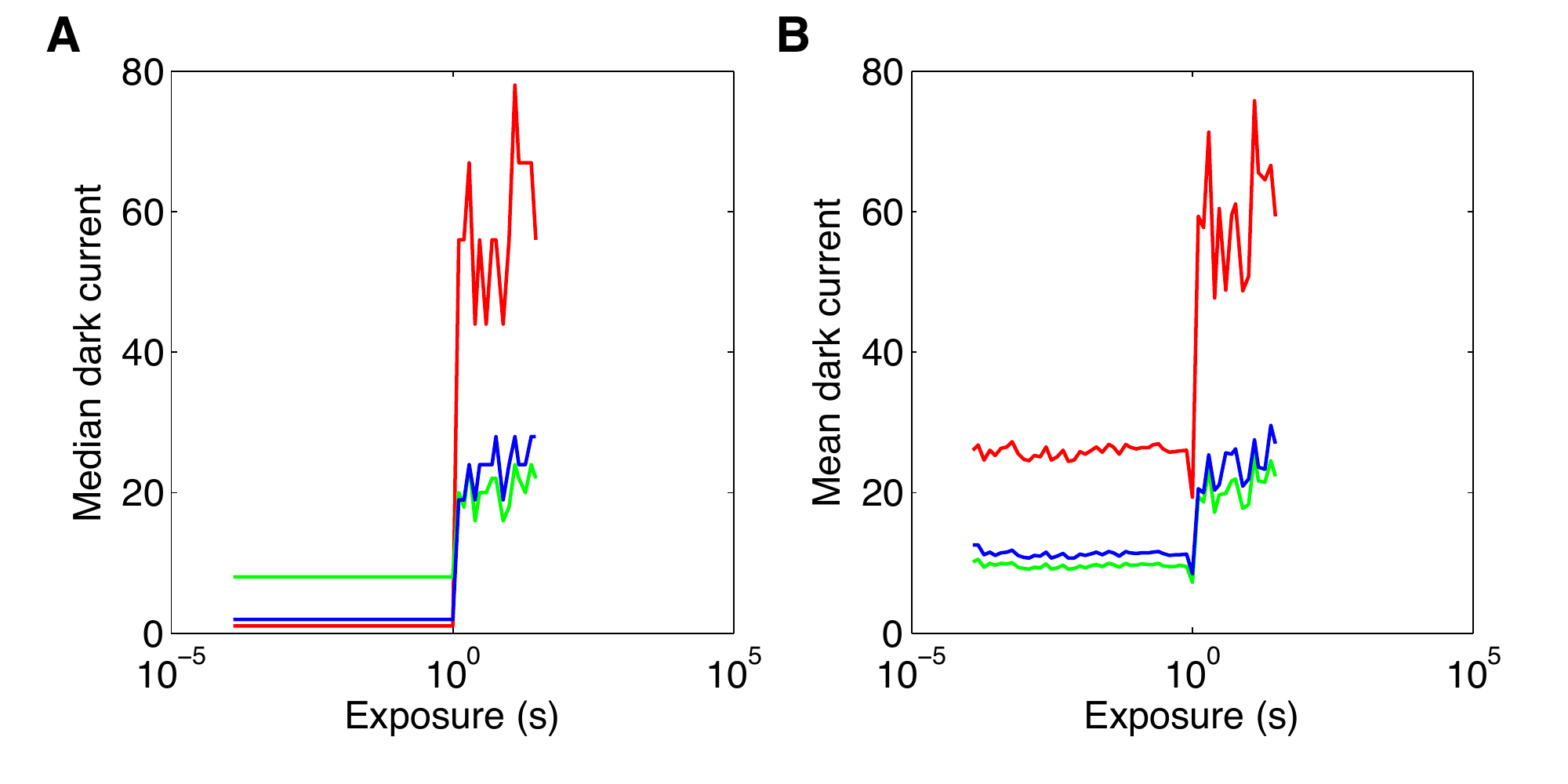}
   \caption{{\bf Dark response by color channel.} {\bf A)} Dark response used for dark subtraction during image processing. For image exposure times below or equal to $1\e{s}$, the dark response for a given color channel (red, green, blue; plot colors correspond to the three color channels) was taken as the median over all the pixels of that color channel  and over all dark image exposure times below $1\e{s}$; for image exposures above $1\e{s}$, we use the median over all the pixels of the same color channel at the given dark image exposure time. {\bf B)} The mean value of dark response across all pixels of the image that are not \emph{``hot''} (i.e. pixels with raw values $<200$, more than $99\%$ of pixels in each color channel), for each color channel, as a function of dark image exposure time. For all dark images, the camera was kept in a dark room with a lens cap on, with the aperture set to minimum ($f22$), and ISO set to 400.} 
 \label{f-dark}
\end{figure}

\emph{Dark subtraction.} Digital cameras typically respond with positive sensor values even when there is no light input (i.e. when an image is acquired with an opaque lens cap in place). This dark response can vary between color channels, with exposure duration, with ISO and with temperature. We did not systematically explore the effect of all these parameters, but we did collect dark images as a function of exposure time for ISO 400 in an indoor laboratory environment. Dark image exposures below 1s generated very small dark response, with the exception of $<0.1\%$ of ``hot pixels'' that  had high (raw value $>200$) response. The median dark response below 1s exposure is less than 11 raw units for all three color channels. For dark image exposures above 1s, the dark response jumped to $\sim 20$ for G and B channels to  and $\sim 56$ for the R channel. The dynamic range of the camera in each color channel is approximately $0-16$k raw units. Dark response values were subtracted from image raw response values as a part of our image preprocessing: for all exposures below or equal to 1s, red dark response value was taken as 1, blue dark response value was taken as 2, and green dark response value was taken as 8 raw units (these values correspond to the mean over exposure durations of the median values across pixels for exposures of less than or equal to 1s); for exposures above 1s, the measured median dark response values across the image frame, computed separately for each exposure, were subtracted. Raw pixel values that after dark subtraction yielded negative values were set to 0. Figure~\ref{f-dark}A shows the dark response that was used for subtraction. Figure~\ref{f-dark}B shows the mean response, excluding the ``hot'' pixels, for comparison. All measurements reported below are for RGB values after dark subtraction. The dark response table is available as a supplementary material to this paper.

\begin{figure}[ht]
\centering
\includegraphics[height=2in ]{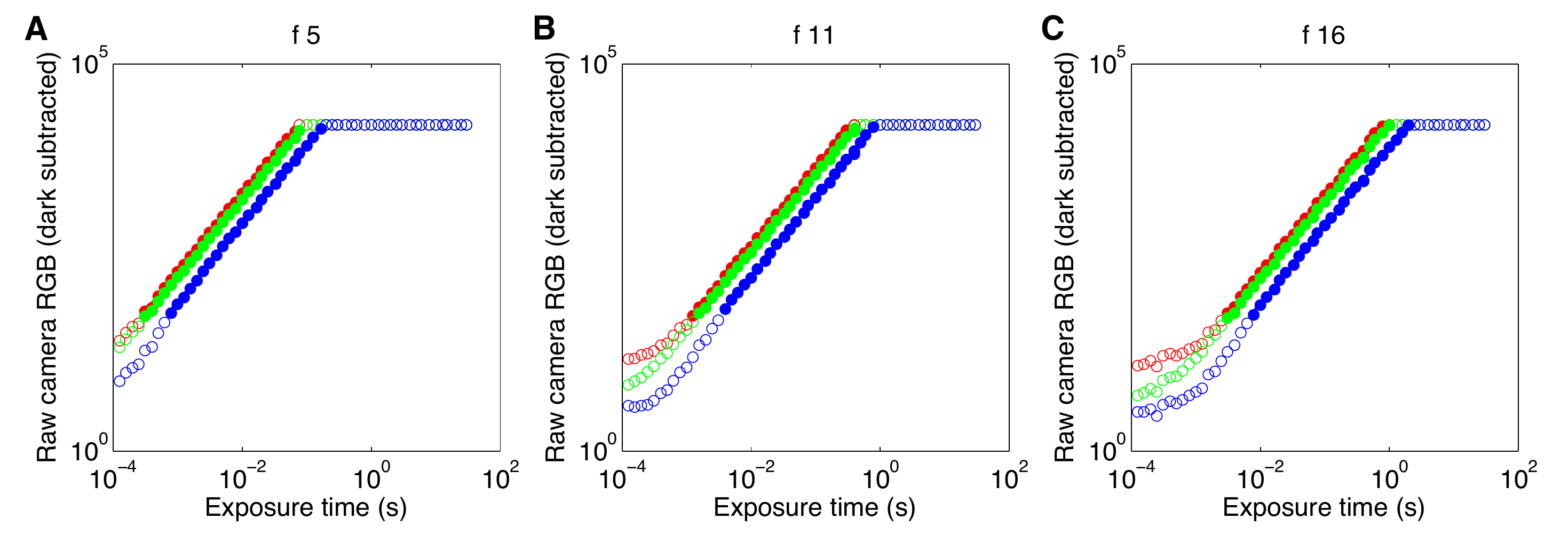}
\caption{{\bf Linearity of the camera in exposure time.} The mean raw RGB response after dark subtraction of the three color channels (red, green, blue; shown in corresponding colors) is plotted against the exposure time in seconds. The values are extracted from images of a white test standard at $f5$ {\bf (A)}, $f11$ {\bf (B)} and $f16$ {\bf (C)} and ISO 200 settings. Full plot symbols indicate raw dark subtracted values between 50 and 16100 raw units; these data points were used to fit linear slopes to each color channel and aperture separately. The fit slopes are 1.01 (R), 1.00 (G), 1.02 (B) for $f5$; 1.00, 0.99, 1.02 for $f11$, and 1.01, 1.00, 1.02 for $f16$.}
\label{f-exposure}
\end{figure}

\emph{Response linearity.} Fundamental to digital camera calibration is a full description of how image values obtained from the camera relate to the intensity of the light incident on the sensors. To measure the D70's intensity-response function, a white test standard was placed $168\e{cm}$ from the camera. The aperture was held constant at $f5, f11$ and $f16$. One picture was taken for each of the 55 possible exposure durations, which ranged from $1/8000 \e{s}$ to $30\e{s}$. Response values were obtained by extracting and averaging RGB values from a $100 \times 100$ pixel region of the image corresponding to light from the white standard.  The camera saturated in at least one channel for longer exposure durations. The camera responses were linear over the duration range $1/320\e{s}$ to $1/3\e{s}$ for the test light level and all apertures used, and for aperture $f5$ approximate linearity extended down to smallest available exposure times, as shown in Fig.~\ref{f-exposure}.  Deviations from linearity are seen at short exposure durations for $f11$ and $f16$.  These presumably reflect a nonlinearity in the sensor intensity-response function at very low dark subtracted response values.

\begin{figure}
\centering
\includegraphics[height=2.5in]{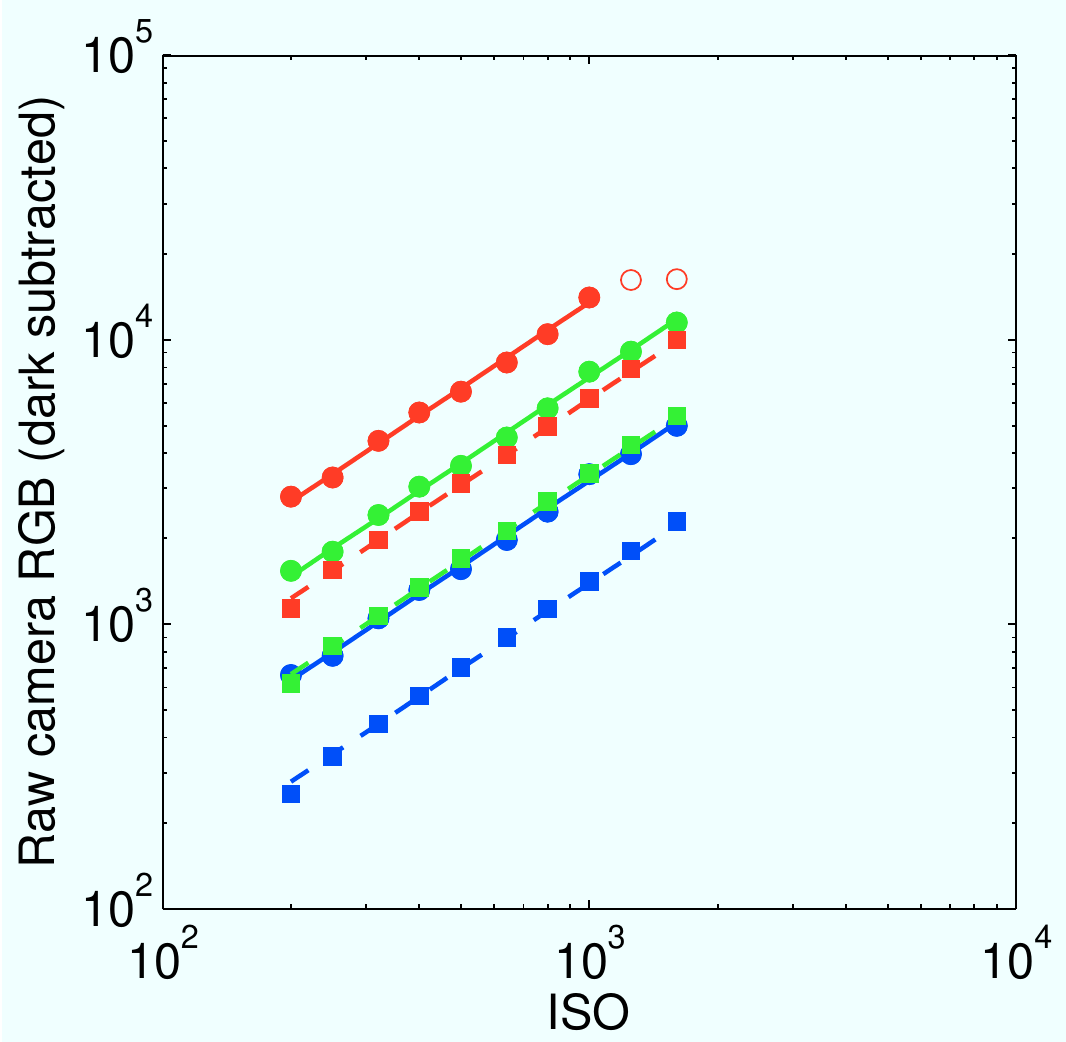}
\caption{{\bf Linearity of the camera in ISO setting.} The mean raw RGB response of the three color channels (red, green, blue; shown in corresponding colors)  is plotted against the ISO setting after dark subtraction, for two values of exposure time (solid line, circles = $1/125\e{s}$; dashed line, squares = $1/250\e{s}$), and $f2.8$ aperture. The lines are linear regressions through non-saturated data points (solid squares or circles; raw dark subtracted values between 50 and 16100); the slopes are 0.99 (R), 0.98 (G), 0.99 (B) for $1/125\e{s}$ exposure and 1.03, 1.02, 1.04 for $1/250\e{s}$ exposure. The camera saturated in the red channel at longer exposure; the corresponding data points (empty red circles) are not included into the linear fit.}
\label{f-iso}
\end{figure}

\emph{ISO linearity.} To examine the effect of the ISO setting on the raw camera response, we acquired a series of 10 images of the white standard with the ISO setting changing from ISO 200 to ISO 1600 at exposures of $1/125\e{s}$ and $1/250\e{s}$. In the regime where the sensors did not saturate, the camera response was linear in ISO for both exposure times, as shown in Fig.~\ref{f-iso}.


\begin{figure}[!ht] 
   \centering
   \includegraphics[height=2in]{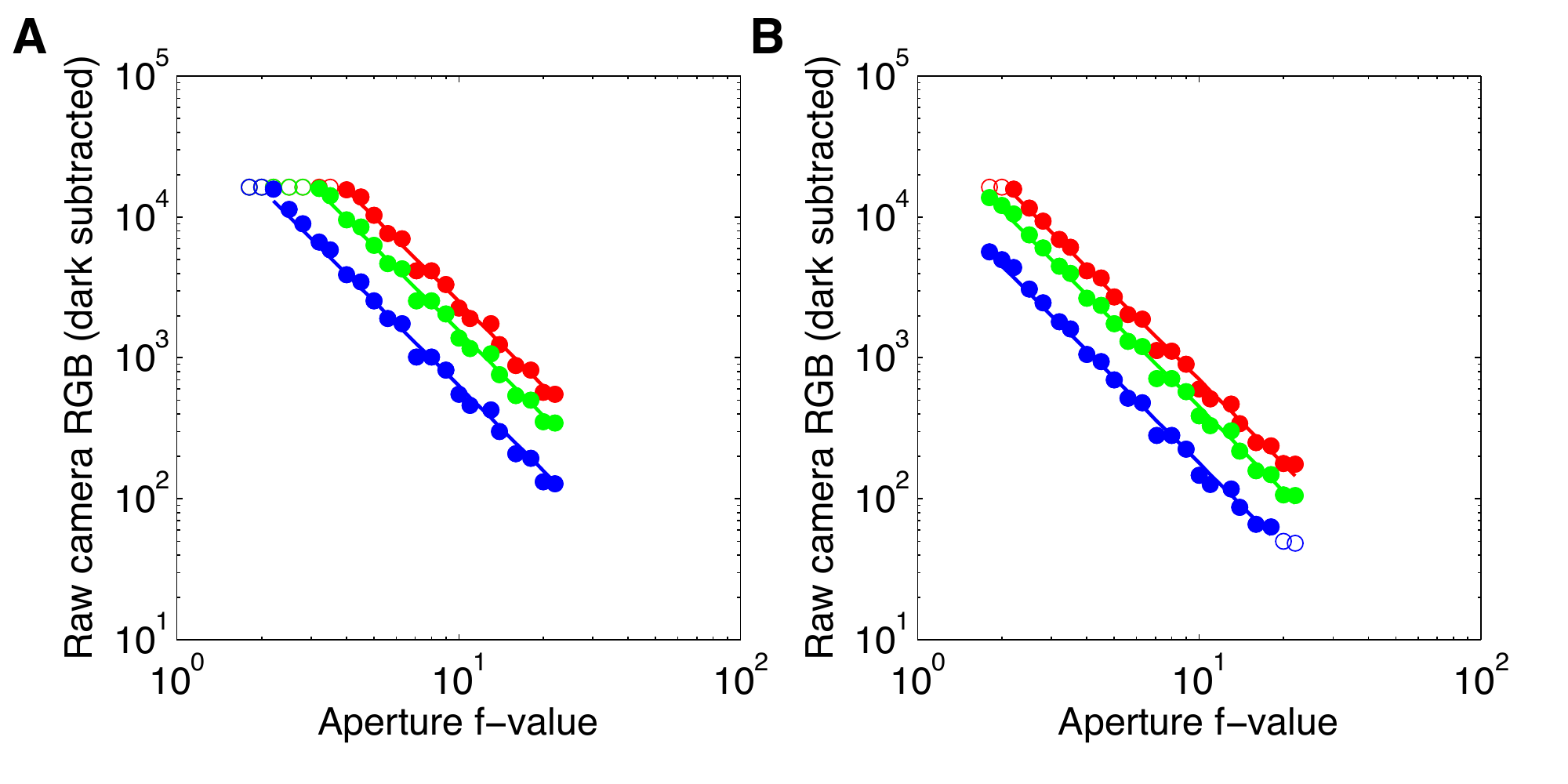}
   \caption{{\bf Camera response as a function of aperture size.} The raw dark subtracted response of the camera exposed to a white test standard {\bf (A)} and a darker secondary image region {\bf (B)}, in three color channels (red, green, blue, shown in corresponding colors), as a function of the aperture (f-value), with exposure held constant to $1/250\e{s}$ and ISO set to 1000. In the regime where the sensors are not saturated and responses are not very small (solid circles, raw dark subtracted values between 50 and 16100), the lines show a linear fit on a double logarithmic scale constrained to have a slope of $-2$ (i.e. $y=kx^{-2}$). Leaving the slopes as free fit parameters yields slopes of $-2.02$ (R), $-2.04$ (G), $-2.10$ (B) for the primary region (white standard) in panel A, and $-2.00$ (R), $-2.03$ (G), $-2.06$ (B) for the secondary region in panel B.  Data points in the saturated or low response regime (empty circles) were not used in the fit.}   The maximum absolute log base 10 deviation of the measurements from the fit lines is 0.1.
 \label{f-aperture}
\end{figure}

\emph{Aperture test.} The aperture size (f-number) of a lens affects the amount of light reaching the camera's sensors.  For the same light source, the intensity per unit area for f-number $x$, $I(x)$, relative to the intensity per unit area for f-number $y$, $I(y)$ should be given by: $I(x) = (y/x)^2 I(y)$. We tested this by measuring the sensor response as a function of exposure duration for all color channels at three different aperture sizes ($f5$, $f11$, and $f16$).  We measured the sensor responses to the white test standard at various aperture sizes to confirm that the response was inversely proportional to the square of the f-number. The camera was positioned facing the white test standard illuminated by a slide projector with a tungsten bulb in an otherwise dark room. The camera exposure time was held at $1/250\e{s}$ and the ISO setting was $1000$.  Images of the white test standard (primary image region) were taken, one for each aperture setting between $f1.8$ and $f22$.  Because image values were saturated at the largest apertures ($f1.8$ and $f2$), we also extracted and analyzed image values from a less intense region in the same image series (secondary image region).  The measurements, shown in Fig.~\ref{f-aperture}A for the primary region and Fig.~\ref{f-aperture}B for the secondary region, confirm that the aperture is operating correctly. There is some scatter of the measured points around the theoretical lines. This may be do to mechanical imprecision for each aperture. We did not pursue this effect in detail, nor attempt to correct for it. Nor did we investigate whether the slightly steeper slopes found for the blue channel represent a slight systematic deviation from expectations for the responses of that channel.

\emph{Standardized RGB values.}
After having verified that the sensor response scales linearly with ISO setting and the exposure, and as $f^{-2}$ with the aperture number $f$ across most of the camera's dynamic range, we define the \emph{standardized RGB values} as dark-subtracted raw camera RGB values, scaled to the reference value of ISO 1000, reference aperture of $f1.8$ and reference exposure time of $1\e{s}$:

\begin{equation}
\mathrm{standardized\_RGB} =(\mathrm{raw\_RGB}-\mathrm{dark\_response})\times\left( \frac{1000}{\mathrm{ISO}}\right) \times \left(\frac{f}{1.8}\right)^2\times \left(\frac{1\mathrm{s}}{\mathrm{exposure}}\right) \label{standardRGB}
\end{equation}

It is the standardized RGB values that provide estimates of the light incident on the camera across our image database.  Our calibration measurements indicate that these values provide good estimates over the multiple decades of light intensity encountered in natural viewing.  Scatter of measurements around the fit lines  shown in the figures above does, however, introduce uncertainty of a few tenths of a log (base 10) unit in the intensity estimates on a finer intensity scale.  Within image, optical factors are likely to introduce systematic variation in sensitivity from the center of the image to the edge, an effect that we did not characterize or correct for.

\begin{figure}[ht]
\centering
\includegraphics[height=2.1in ]{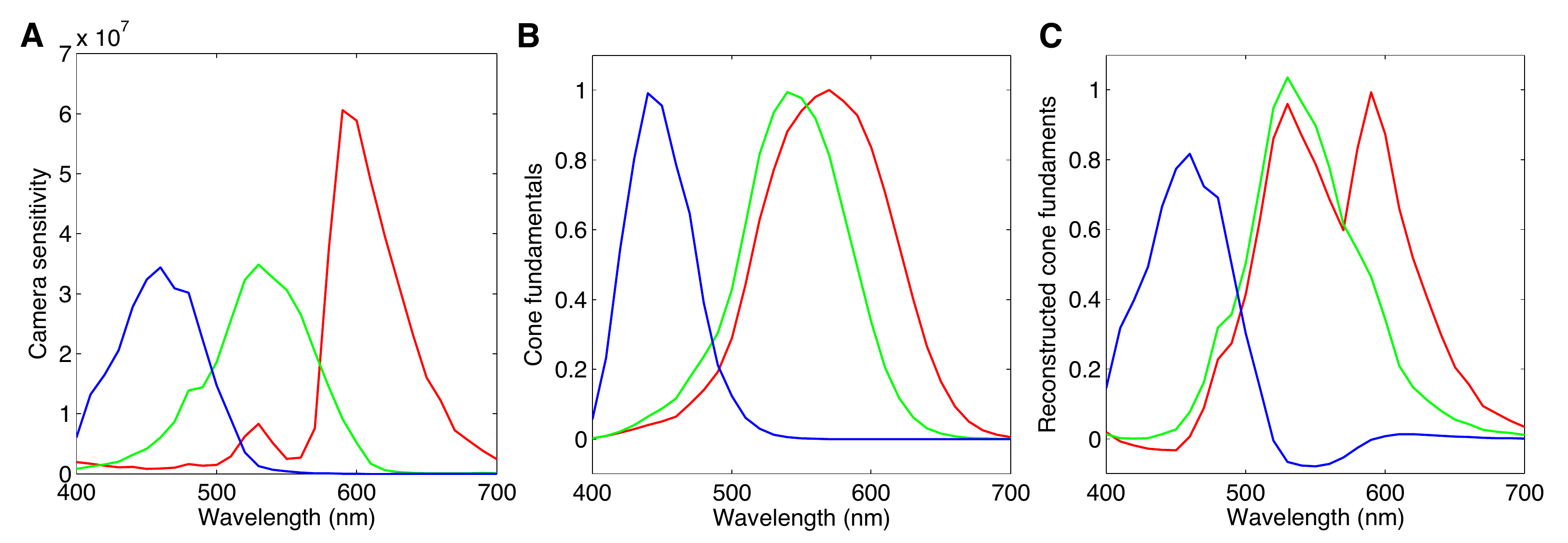}
\caption{{\bf Spectral response of the camera.} {\bf A)} The spectral sensitivity curves plotted here convert spectral radiance into standardized camera RGB values. {\bf B)} The LMS cone fundamentals \cite{Sharpe:Stockman:2005,CIE2007} for L (red), M (green) and S (blue) cones. Note that the fundamentals are normalized to have a maximum of 1. {\bf C)} A linear transformation can be found that transforms R,G,B readings from the camera with sensitivities plotted in (A) into reconstructed fundamentals L'M'S' shown here, such that L'M'S' fundamentals are as close as possible (in mean-squared-error sense) to the true LMS fundamentals shown in (B).  }
\label{f-spectra}
\end{figure}

\emph{Spectral response.} Next, we measured the spectral sensitivities of the camera sensors. A slide projector (Kodak Carousel 440 ~\cite{KodakCarousel}), the Nikon D70 Camera, and a spectroradiometer~\cite{PhotoResearch} were positioned in front of the white test standard. Light from the projector was passed through one of 31 monochromatic filters, which transmitted narrowband light between $400\e{nm}$ and $700\e{nm}$, at $10\e{nm}$ intervals. For each filter, a digital picture ($f1.8$, ISO $1000$, and varying exposure duration) and a spectroradiometer reading were taken.  The RGB data were dark subtracted and converted to standardized RGB values, and then compared to the radiant power read by the spectroradiometer to estimate the three sensor's spectral sensitivities.  For this purpose, we summed power over wavelength and treated it as concentrated at the nominal center wavelength of each narrowband filter.  For these images, dark subtraction was performed with dark images acquired at the same time as the spectral response images were acquired.  These dark images were taken by occluding the light projected onto the white test standard, so that they accounted for any stray broadband light in the room as well as for sensor dark noise.  The measured spectral sensitivities are shown in Fig.~\ref{f-spectra}.  These may be used to generate predictions of the camera sensor response to arbitrary spectral light sources.  To compute predicted standardized RGB values, the input spectral radiance measured in units of $\e{W/(nm\cdot sr\cdot m^2)}$ should be integrated against each spectral sensitivity.

\begin{figure}
\centering
\includegraphics[height=2.0in]{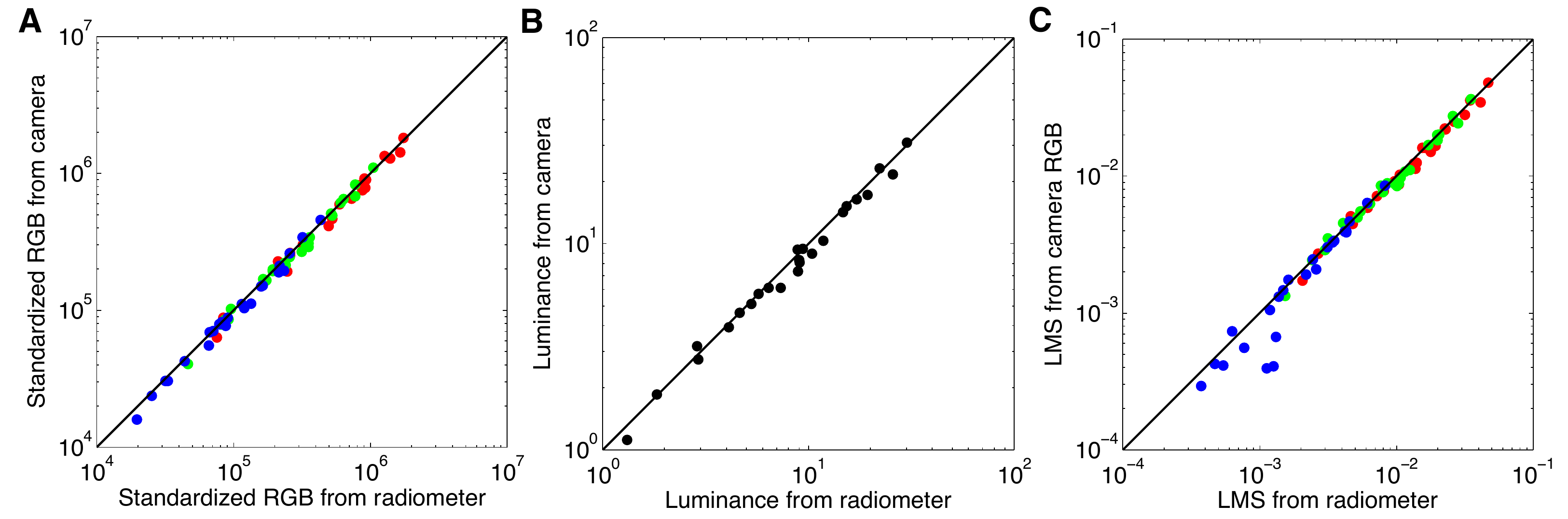}
\caption{{\bf Checking the camera calibration.} Digital images and direct measurements of spectral radiance were obtained for the 24 color swatches of the Macbeth color checker chart. {\bf A)} Raw standardized RGB values were obtained from the camera images as described in \emph{Materials and Methods}. RGB response was also estimated directly from the radiometric readings via the camera spectral sensitivities shown in Fig.~\ref{f-spectra}A. Plotted is the comparison of the corresponding $24\times 3$ RGB values; black line denotes equality. {\bf B)} The luminance in $\mathrm{cd/m^2}$ measured directly by the radiometer compared to the luminance values obtained from the standardized camera RGB values. {\bf C)}  This plot shows the correspondence between the Stockman-Sharpe/CIE 2-degree LMS cone coordinates estimated from the camera and those obtained from the measured spectra. Plot symbols red, green and blue indicate L,M,S values respectively, and the data are for the 24 MCC squares.}
\label{f-mcc}
\end{figure}

To check the end-to-end accuracy of our camera RGB calibration, we acquired an image of the Macbeth color checker chart, extracted the raw RGB values for each of its 24 swatches, and converted these to the standardized RGB representation.  We then compared these measured values against predictions obtained from direct measurments of the spectral radiance reflected from each swatch.  The comparison in Fig.~\ref{f-mcc}A shows an excellent fit between predicted and measured values.


\begin{figure}[!ht] 
   \centering
   \includegraphics[width=2.5in]{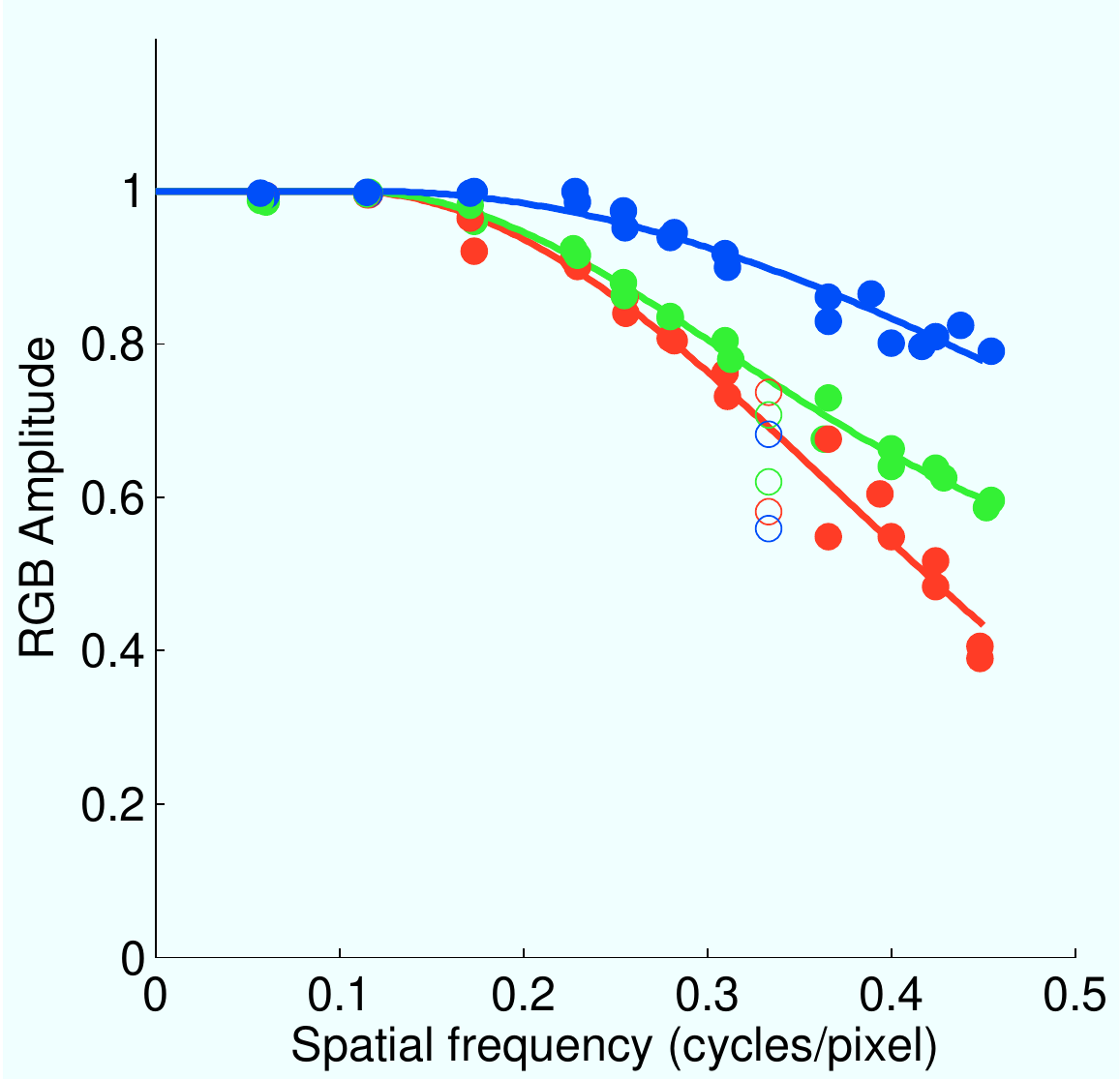}
   \caption{{\bf Camera spatial MTF.}  Estimated MTF is plotted as a function of spatial frequency for the red, green, and blue image planes (shown in corresponding colors). Solid lines show empirical fits to $y=a\exp\left\{-b(x-d)^2\right\} + (1-a) \exp(-cx^2)$, where for all $x\leq d$, $y$ is set to 1 and where any fit values $y$ greater than 1 were also set to 1. The fit parameters are  $a = 1.00, b = 6.96, c = 0.50, d = 0.10$ for red channel, $a = 0.52, b = 12.37, c = 0.02, d = 0.11$ for green channel, and  $a = 1.00, b = 2.24, c = 0.50, d = 0.12$ for blue channel.  MTF values at $\sim 0.33$ cycles/pixel (empty plot symbols) systematically deviated from the rest and were excluded from the fit.} 
 \label{MTF}
\end{figure}

\emph{Spatial Modulation Transfer Function}. The modulation transfer function (MTF) describes how well contrast information is transmitted through an optical system.   We estimated the MTF of our camera for the R, G, and B image planes individually, to compare the transmission of contrast information at various spatial frequencies for light intensity measured in each of the three types of camera photosensors.
To do this, we first imaged a high-contrast black and white square-wave grating at a set of distances relative to the camera and extracted a horizontal grating patch from each image.  This was summed over columns to produce a one-dimensional signal for each image plane.  We then used the fast-fourier transform to find the amplitude and spatial frequency of the fundamental component of the square wave-grating.  In this procedure, we tried various croppings of the grating and chose the one that yielded maximum amplitude at the fundamental.  This minimized spread of energy in the frequency domain caused by spatial sampling.  We then reconstructed a one-dimensional spatial domain image by filtering out all frequencies except for the fundamental, and computed the image contrast as the difference between minimum and maximum intensity divided by the sum of minimum and maximum intensity.  Figure~\ref{MTF} plots image contrast as a function of spatial frequency (cycles / pixel) for each color channel.   At high spatial frequencies, the blue channel is blurred least and the red channel is blurred most, presumably due to chromatic aberration in the lens.  We did not use these data in our image processing chain, but fits to the data are available (see caption of Fig.~\ref{MTF}) either to correct for camera blurring for applications where that is desirable or to estimate the effect of camera blur on any particular image analysis.  It should be noted that the MTF will depend on a number of factors, including f-stop, exactly how the image is focussed, and the exact spectral composition of the incident light.  The measurements were made for $f5.6$ and the camera's auto focus procedure.  In addition, the measurements do not account for affects of lateral chromatic aberrations, which can produce magnification differences across the images seen by the three color channels.  For these reasons, the MTF data should be viewed as an approximation.

\begin{figure}[!ht] 
   \centering
   \includegraphics[width=2.5in]{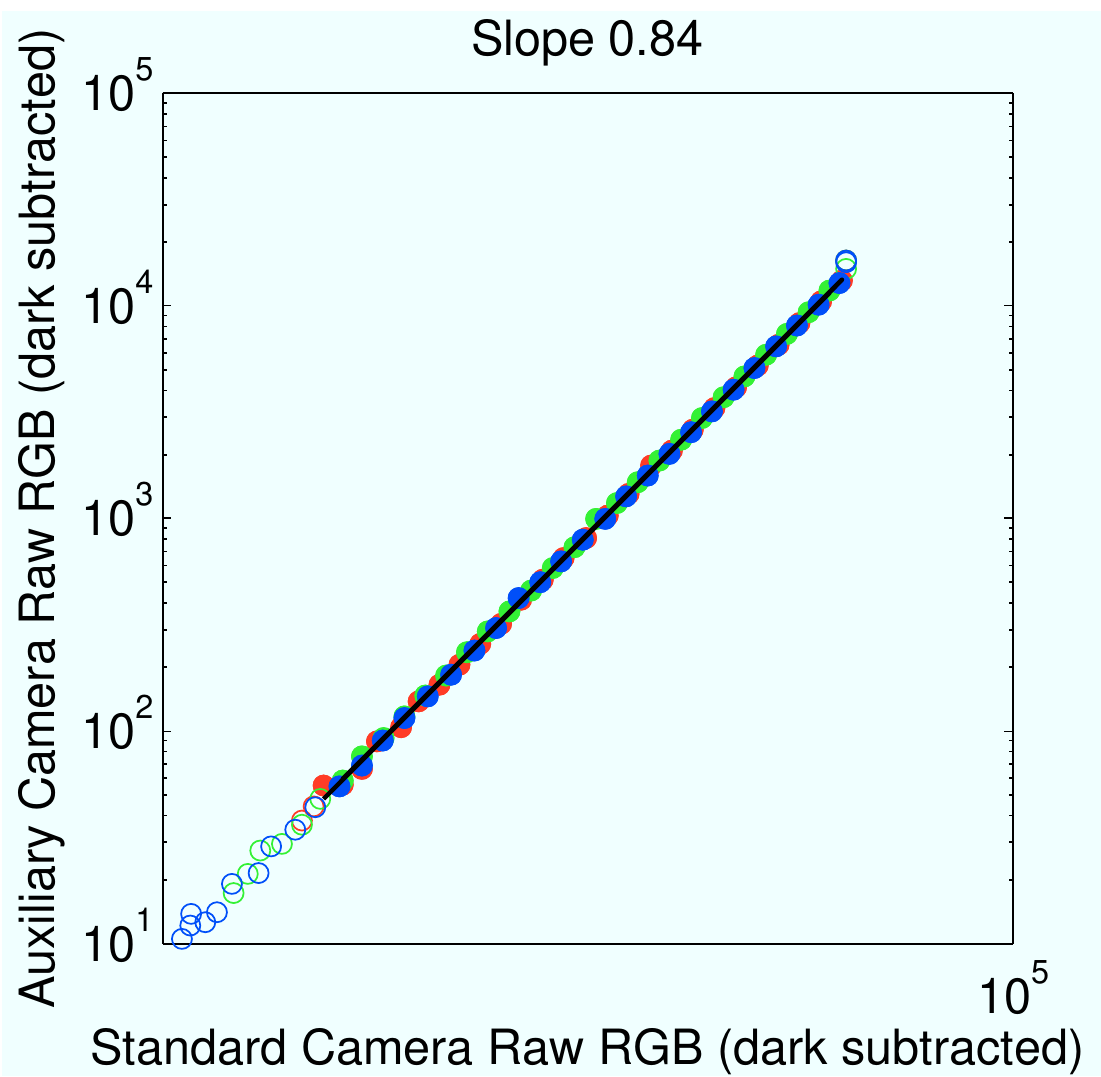}
   \caption{{\bf Comparison between standard and auxiliary camera.} Raw dark subtracted values for  three color channels (red, green, blue; shown in corresponding colors) of the same standard taken at different exposures using the standard and auxiliary cameras. Black like is a linear fit on non-saturated points (raw dark subtracted pixel values between 50 and 16100), with fit slope of 0.84. Images were acquired at $f5.6$ and ISO $400$.  Similar slopes ($0.8\pm 0.05$) were found when positions of cameras imaging the white standard were slightly changed (see text), and when the camera readouts were compared on color swatches of the Macbeth color checker ($f2.8$, ISO $400$.) } 
 \label{cc}
\end{figure}

\emph{Camera comparisons.}  We compared the response properties of two Nikon D70 cameras -- the standard camera (used for the analysis detailed above), and a second auxiliary camera. We found that, under identical conditions, the sensor response of the auxiliary camera was approximately $80\%\pm 5\%$ of the response of the standard camera, across all channels, response times, and apertures (see e.g. Fig.~\ref{cc}). We compared the two cameras imaging the white test standard with varying exposure durations, and for color swatches of the Macbeth color checker. We found that the ratio of camera responses was relatively constant throughout the linear range of exposure durations. To ensure that the response difference was not a result of the slightly different positions of the tripods on which each camera was mounted, we also took pictures of the white standard with the cameras on the same tripod (sequentially), then with the auxiliary camera $25\e{cm}$ to the left of the standard camera and visa versa. The difference in camera responses could not be accounted for by small positional differences. Therefore, the different camera response magnitudes probably indicate a characteristic of the cameras themselves. At very fast or very slow exposure durations, the responses of both cameras were dominated by dark-response and sensor saturation, respectively, and the 80\% response ratio did not apply. This analysis suggested that a complete calibration of the auxiliary camera was unnecessary. Instead, the spectral sensitivities of R, G, B sensors for the auxiliary camera were defined to be 0.80 times the spectral sensitivities of the standard camera; by using this simple multiplicative conversion to bring the auxiliary and standard cameras into accordance, the remaining image transformation steps remain unchanged between both cameras.

{\bf Colorimetry.}  For applications to human vision, it is useful to convert the camera RGB representation to one that characterizes how the incident light is encoded by the human visual system.  For this reason, we used the camera data to estimate the photopigment isomerization rates of the human L, M, and S cones produced by the incident light at each pixel.  Because the camera spectral sensitivities are not a linear transformation of the human cone spectral sensitivities, the estimates will necessarily contain some error~\cite{horn1984}, and different techniques for performing the estimation will perform better for some ensembles of input spectra than for others~\cite{Wandell1987,finlayson1996,zhang2004}.  Here we employed a simple estimation method.  We first found the linear transformation of the camera's spectral sensitivities that provided the best least-squares approximation to the Stockman-Sharpe/CIE 2-degree (foveal) estimates of the cone fundamentals~\cite{Sharpe:Stockman:2005,CIE2007}.  Figure~\ref{f-spectra}B plots the cone fundamentals and Fig.~\ref{f-spectra}C plots the approximation to these fundamentals obtained from the camera spectral sensitivities.  By applying this same transformation to the standardized RGB values in an image, we obtain estimates of the LMS cone coordinates of the incident light.  For each cone class, these are proportional to the photopigment isomerization rate for that cone class.

To determine the constant of proportionality required to obtain LMS isomerization rates from the LMS cone coordinates for each cone class, we estimated photopigment isomerization rates directly from the measured spectral radiance of the light reflected from the 24 color swatches of the Macbeth color checker.  This was done using software available as part of the Psychophysics Toolbox~\cite{Brainard:1997,Yin2006} along with the parameter values for human vision provided in Table~\ref{t-3}.  We also computed the response of the LMS cones from these same measured spectra using the cone response functions plotted in Fig.~\ref{f-spectra}B.  For each cone class, we then regressed the 24 isomerization rates against the 24 corresponding cone coordinates to obtain the scalar required to transform cone coordinate to isomerization rate.  For the L, M, and S cones respectively, the scalar values are $1.75\E{5}, 1.60\E{5}, 3.49\E{4}$, with units that yield isomerizations per human cone per second.

For certain analyses, it is useful to provide grayscale versions of the images.  We  thus computed estimates of the luminance $Y$ (in units of candelas per meter squared with respect to the CIE 2007 two-degree specification for photopic luminance spectral sensitivity~\cite{CIE2007}).    These were obtained as a weighted sum of the estimated Stockman-Sharpe/CIE 2-degree LMS cone coordinates: $Y=433.94L+275.82M-0.09S$; these weights yield luminance in units of $\e{cd/m^2}$.

Figure~\ref{f-mcc}B shows the comparison between the luminance for the Macbeth color checker the the luminance measured directly using the radiometer.  Figure~\ref{f-mcc}C shows the comparison between the Stockman-Sharpe/CIE 2-degree LMS cone coordinates estimated from the camera and those computed directly from the radiometrically measured spectra.   Overall the agreement is good, with the exception of the S-cone coordinates for a few color checker squares.  These deviations occur because the camera spectral sensitivities are not an exact linear transformation away from the cone fundamentals.

{\bf Image extraction and data formats.}  For each image in every database album (folder), the following operations are carried out:
\begin{enumerate}
\item {\bf Raw image .NEF to .PPM conversion.}  Images with name pattern {\tt DSC\_\#\#\#\#.NEF} (where \# are image serial numbers) were first extracted from the camera as proprietary Nikon Electronic Format (NEF) files (approximately 6Mb in size), which record ``raw'' sensor values. These files were converted to PPM files using {\sc dcraw v5.71} \cite{dcraw} in Document Mode (no color interpolation between RGB sensors) by using the {\em -d} flag, and written out as 48 bits-per-pixel (16 bits per color channel) PPM files by using the \emph{-4} flag; we note that the behavior of {\sc dcraw} is highly version dependent. In addition, we extracted the following image meta-data from the NEF file: the camera serial number (to determine whether an image was taken using the \emph{standard} or \emph{auxilliary} camera), exposure duration, f-value, and the ISO setting. We also extracted JPG formatted images of the NEF images; both JPG and NEF formats are available from the database as {\tt DSC\_\#\#\#\#.NEF} and {\tt DSC\_\#\#\#\#.JPG}.

\item {\bf PPM to raw RGB conversion.} PPM images were loaded by our Matlab script as $2014\times 3038$ matrices. Because of the CCD mosaic, pixels in this matrix that have no corresponding CCD sensor contain zeros.  Subsequently, each color plane of the image was block-averaged in $2\times 2$ blocks into a resulting $1007\times 1519\times 3$ raw RGB format, where the last dimension indicates the color channel ($R=1,G=2,B=3$); red and blue pixels are weighted by 4 and green pixels by 2, reflecting the number of sensors for each color in every $2 \times 2$ block. The resulting matrix has raw units spanning the range from 0 to 16384 for red and blue channels, and from 0 to 16380 for the green channel. It is available as a Matlab matrix with name pattern {\tt DSC\_\#\#\#\#\_RGB.mat} from the image database.
\item {\bf Raw RGB to standardized, dark-subtracted RGB.} We next took the resulting block-averaged RGB images and subtracted the dark response, characteristic for each color channel and for the image exposure duration, using the dark response values shown in Fig.~\ref{f-dark}A. After subtracting the dark response, each image was multiplied by the scale factor, to yield a RGB dark-subtracted image standardized to one second exposure duration, ISO 1000 setting, and f-value of 1.8, as prescribed by Eq~(\ref{standardRGB}).

\item {\bf Standardized, dark-subtracted RGB to LMS and Luminance formats.} By regressing R,G,B camera sensitivities (measured in Fig.~\ref{f-spectra}) against known L,M,S cone sensitivities, we obtain a $3\times 3$ matrix that approximately transforms from the RGB into the LMS color coordinate system. Each pixel of the image can thus be transformed into the LMS system. By multiplying the image in the LMS system with the L,M,S isomerisation rate factors (see Colorimetry above), each image is now expressed in units of L,M,S isomerizations per second and saved as a Matlab  $1007\times 1519\times 3$ matrix with name pattern {\tt DSC\_\#\#\#\#\_LMS.mat}. In parallel, the image in the LMS system can be transformed into a grayscale image with pixels in units of cd/m$^2$, by summing over the three (L, M, S) color channels of each pixel with appropriate weights (see Colorimetry above). The grayscale image is saved as a $1007\times 1519$ matrix with name pattern {\tt DSC\_\#\#\#\#\_LUM.mat}.

\item {\bf Image metadata in AUX files.} For each image, the database contains a small auxiliary Matlab structure saved as {\tt DSC\_\#\#\#\#\_AUX.mat} that contains image meta-data: {\bf (i)} the EXIF fields extracted from the JPG and NEF images, among others, the aperture, exposure settings, ISO sensitivity, camera serial number, and timestamp of the image; {\bf (ii)} the identifier of the image in the master database (the image name and album), {\bf (iii)} the timestamp when the image processing was done, {\bf (iv)} fraction of pixels at saturation in the RGB image, {\bf (v)} a warning flag indicating whether the exposure time is out of the linear range or if too many pixels are either at low (dark-response) intensities or at saturation, and {\bf (vi)}, several human-assigned annotations (not available for all images), e.g. the distance to the target, the tripod setting, qualitative time of day, lighting conditions etc. Most images are taken in normal, daylight viewing conditions, and should therefore have no warning flags; some images, however, especially ones that are part of the exposure or time-of-day series, can have saturated or dark pixels, and it is up to the database users to properly handle such images.

\end{enumerate}

\section{Discussion}

We have collected a large variety of images that include many snapshots of what a human observer might conceivably look at, such as images of the horizon and detailed images of the ground, trees, bushes, and baboons. This riverine / savanna environment in Okavango delta was chosen because it is similar to where human eye is thought to have evolved. The images were taken at various times of the day, and with various distances between the camera and the objects in the scene. A subset of albums focuses on particular objects, such as berries and other edible items, in their natural context.  We also provide close-up images of natural objects taken from different distances and accompanied by a ruler (from which an absolute scale of the objects can be inferred). Other albums can be used to explore systematic variation of the image statistics with respect to a controllable parameter, for instance, the variation in the texture of sand viewed from various distances, the change in color composition of the sky viewed at regular time intervals during the day, or the same scene viewed at an increasing angle of the camera from the vertical, from pointing towards the ground to pointing vertically at the sky. A large fraction of the database contains images of various types of Botswana scenery (flood plains, sand plains, woods, grass) sampled without purposely focusing on any particular object.

Many studies of biological visual systems  explore the hypothesis that certain measurable properties of neural visual processing systems reflect optimal adaptations to the structure of natural visual environments~\cite{Barlow:1982,Attick:Redlich:1992,Maloney:1986,Regan:Julliot:2001,Lythgoe:Partridge:1989,Chittka:Menzel:1992,Osorio:Vorobyev:1996,Lewis:Zhaoping:2006,Tkacik2010,Ratliff2010,Garrigan2010}.  The images we have collected allow for a reliable estimation of the properties of natural scenes, such as local luminance histograms, spatial two-point and higher-order correlation functions, scale invariance, color and texture content, and the statistics of oriented edges (e.g., colinearity and cocircularity)~\cite{Ruderman:1994p258,Ruderman:1994p1353,Ruderman:1997,Mante:2005p1214,Sigman:2001p30,Karklin:2003p1177} (see also a topical issue on natural systems analysis \cite{Geisler2009}).  Currently, we include only images from a single environment, which allows for adequate sampling of the relevant statistical features that distinguish these natural scenes from random ones.  In the future, we plan to expand the dataset to include images of other natural, as well as urban, environments, some of which have already been acquired.  This expanded image set will allow for sampling of the statistical features that distinguish different environments.

\section*{Acknowledgements}
We thank E. Kanematsu for helpful discussions. This research was supported by grant NEI R01 EY10016 (DHB) and by grants NSF IBN-0344678 (VB), NSF EF-0928048 (VB), NIH R01 EY08124 (VB, PS).

\begin{table}[!ht]
\caption{
{\bf Albums 01a--61a of the Botswana dataset.}}
\footnotesize
\centering
\begin{tabular}{|c|c|c|}
\hline
Album & Keywords & Tags 	\\ \hline\hline
{\tt cd01a} & baboons, grass, tress, bushes &	\\ \hline
{\tt cd02a} & baboons, grass, tress, bushes &   \\ \hline
{\tt cd03a} & trees	& sequence: vertical angle \\ \hline
{\tt cd04a} & trees, grass & sequence: aperture \\ \hline
{\tt cd05a} & trees, leaves & sequence: aperture \\ \hline
{\tt cd06a} & flood plain, grass, trees, sky & \\ \hline
{\tt cd07a} & baboons, horizon & sequence: aperture \\ \hline
{\tt cd08a} & sand plain, horizon & sequence: aperture \\ \hline
{\tt cd09a} & sand plain, bushes, horizon & \\ \hline
{\tt cd10a} & bushes, trees & sequence: aperture \\ \hline
{\tt cd11a} & forrest & sequence: aperture \\ \hline
{\tt cd12a} & forrest, ground & \\ \hline
{\tt cd13a} & flood plain, water, horizon & \\ \hline
{\tt cd14a} & horizon, sunset & sequence: time, short+long exposure \\ \hline
{\tt cd15a} & leaves, grass, ground & \\ \hline
{\tt cd16a} & road, cars, buildings, town & \\ \hline
{\tt cd17a} & road, cars, buildings, town & \\ \hline
{\tt cd18a} & tree, ground & sequence: aperture \\ \hline
{\tt cd19a} & sand plain, ground & closeup \\ \hline
{\tt cd20a} & sand plain, horizon & sequence: aperture \\ \hline
{\tt cd21a} & sand plain, sand ground, cloudy sky & \\ \hline
{\tt cd22a} & baboons, sand ground, horizon & sequence: aperture \\ \hline
{\tt cd23a} & leaves, bushes, ground, sky & \\ \hline
{\tt cd24a} & leaves & sequence: aperture \\ \hline
{\tt cd25a} & plant, leaves & closeup \\ \hline
{\tt cd26a} & leaves, bushes & sequence: aperture \\ \hline
{\tt cd27a} & flood plain, grass, water, horizon & sequence: time, short+long exposure \\ \hline
{\tt cd28a} & termite mound, horizon & sequence: aperture \\ \hline
{\tt cd29a} & sand plain, horizon & sequence: aperture \\ \hline
{\tt cd30a} & baboons, sand plain, bushes, trees, horizon & \\ \hline
{\tt cd31a} & bushes, termite mound & sequence: scale, sequence: vertical angle \\ \hline
{\tt cd32a} & flood plain, water, grass, horizon & \\ \hline
{\tt cd33a} & flood plain, water, grass, horizon  &\\ \hline
{\tt cd34a} & forrest, tree, leaves & sequence: scale \\ \hline
{\tt cd35a} & grass, trees & sequence: scale \\ \hline
{\tt cd36a} & bark & closeup, sequence: aperture \\ \hline
{\tt cd37a} & leaves & sequence: aperture \\ \hline
{\tt cd38a} & sand plain, trees, horizon, ground & \\ \hline
{\tt cd39a} & sand plain, horizon & sequence: aperture \\ \hline
{\tt cd40a} & bark & closeup \\ \hline
{\tt cd41a} & forrest, trees, leaves, ground & \\ \hline
{\tt cd42a} & fruit, nuts & closeup, on table \\ \hline
{\tt cd43a} & grass plain, baboons, water, & \\ \hline
{\tt cd44a} & baboons, grass, human & \\ \hline
{\tt cd45a} & sand plain, rock & sequence: scale \\ \hline
{\tt cd46a} & plain, tree & sequence: scale \\ \hline
{\tt cd47a} & tree trunk, forrest & sequence: scale \\ \hline
{\tt cd48a} & tree trunk, sand plain & sequence: scale \\ \hline
{\tt cd49a} & tree stump, ground & sequence: scale \\ \hline
{\tt cd50a} & fruit, plant & closeup, sequence: aperture \\ \hline
{\tt cd51a} & tree, sky & sequence: aperture \\ \hline
{\tt cd52a} & palm tree & sequence: aperture \\ \hline
{\tt cd53a} & trees, grass, forest & sequence: time (all day) \\ \hline
{\tt cd54a} & sand plain, horizon & sequence: time (all day) \\ \hline
{\tt cd55a} & grass plain, horizon & sequence: time (all day) \\ \hline
{\tt cd56a} & baboons, grass & closeup \\ \hline
{\tt cd57a} & baboons, grass, trees, bushes & \\ \hline
{\tt cd58a} & baboons, grass, trees, horizon & \\ \hline
{\tt cd59a} & baboons, grass, tree stumps & \\ \hline
{\tt cd60a} & horizon & sequence: time (sunset), short+long exposure \\ \hline
{\tt cd61a} & sky, moon & sequence: time (night) \\ \hline
\end{tabular}
\begin{flushleft}Keywords provide a short description of the image content. Tags provide additional information about how the images were acquired. A ``sequence'' tag means that the same scene was taken several times while changing a parameter, e.g., ``sequence: aperture'' means that the scene was photographed while changing the aperture (and simultaneously the exposure time), ``sequence: scale'' means that the object is photographed at decreasing distance to the camera, ``sequence: time'' means that the scene is photographed at approximately equal time intervals, etc.
\end{flushleft}
\label{t-1}
 \end{table}
\begin{table}[!ht]
\caption{
{\bf Albums 01b--35b of the Botswana dataset.}}
\footnotesize
\centering
\begin{tabular}{|c|c|c|}
\hline
Album & Keywords & Tags	\\ \hline\hline
{\tt cd01b} & dirt, ground & closeup, sequencescale: scale, ruler	 \\ \hline
{\tt cd02b} & sand, ground & closeup, sequence: scale, ruler	 \\ \hline
{\tt cd03b} & salt deposits, ground & closeup, sequence: scale, ruler	 \\ \hline
{\tt cd04b} & scrub, ground & closeup, sequence: scale, ruler	 \\ \hline
{\tt cd05b} & sticky grass & closeup, sequence: scale, ruler	 \\ \hline
{\tt cd06b} & marula nut  & closeup, sequence: scale, ruler	 \\ \hline
{\tt cd07b} & sausage fruit  &closeup,  sequence: scale, ruler	 \\ \hline
{\tt cd08b} & elephant dung & closeup, sequence: scale, ruler	 \\ \hline
{\tt cd09b} & old figs & closeup, sequence: scale, ruler	 \\ \hline
{\tt cd10b} & fresh figs & closeup, sequence: scale, ruler	 \\ \hline
{\tt cd11b} & old jackelberry & closeup, sequence: scale, ruler	 \\ \hline
{\tt cd12b} & woods, ground & closeup, sequence: scale, ruler	 \\ \hline
{\tt cd13b} & fresh buffalo dung & closeup, sequence: scale, ruler	 \\ \hline
{\tt cd14b} & fresh jackelberry &closeup,  sequence: scale, ruler	 \\ \hline
{\tt cd15b} & semiold palm nut & closeup, sequence: scale, ruler	 \\ \hline
{\tt cd16b} & old palm nut & closeup, sequence: scale, ruler	 \\ \hline
{\tt cd17b} & fresh palm nut&closeup,  sequence: scale, ruler	 \\ \hline
{\tt cd18b} &  fresh sausage fruit & closeup, sequence: scale, ruler	 \\ \hline
{\tt cd19b} & semifresh sausage fruit & closeup, sequence: scale, ruler	 \\ \hline
{\tt cd20b} & old buffalo dung & closeup, sequence: scale, ruler	 \\ \hline
{\tt cd21b1} & marula tree bark & closeup, sequence: scale, ruler	 \\ \hline
{\tt cd21b2} & palm tree bark & closeup, sequence: scale, ruler	 \\ \hline
{\tt cd22b1} & fig tree bark & closeup, sequence: scale, ruler	 \\ \hline
{\tt cd22b2} & jackelberry tree bark & closeup, sequence: scale, ruler	 \\ \hline
{\tt cd23b} & saussage tree bark & closeup, sequence: scale, ruler	 \\ \hline
{\tt cd24b1} & sage & closeup, sequence: scale, ruler	 \\ \hline
{\tt cd24b2} & termite mound & closeup, sequence: scale, ruler	 \\ \hline
{\tt cd25b} & woods, horizon, sky & sequence: vertical angle	 \\ \hline
{\tt cd26b} & sand plain, horizon, sky & sequence: vertical angle	 \\ \hline
{\tt cd27b} & flood plain, water, horizon, sky & sequence: vertical angle	 \\ \hline
{\tt cd28b} & bush, sky & sequence: vertical angle	 \\ \hline
{\tt cd29b1} & sand plain & sequence: height	 \\ \hline
{\tt cd29b2} &woods & sequence: height	 \\ \hline
{\tt cd30b1} & flood plain  & sequence: height \\ \hline
{\tt cd30b2} & bush & sequence: height \\ \hline
{\tt cd31b} & fresh elephant dung & closeup,  sequence: scale, ruler \\ \hline
{\tt cd32b} & sky, no clouds & sequence: time (all day)	 \\ \hline
{\tt cd33b1} & horizon & sequence: time (sunrise)	 \\ \hline
{\tt cd33b2} & horizon & sequence: time (sunset)	 \\ \hline
{\tt cd34b1} & woods, bushes & sequence: time (sunset) \\ \hline
{\tt cd34b2} & woods, bushes & sequence: time (sunrise)	 \\ \hline
{\tt cd35b} &grass, horizon & sequence: time (all day)	 \\ \hline
\end{tabular}
\begin{flushleft} Albums with the ``ruler'' tag have a green ruler present in the scene so that the absolute size of the objects can be determined.
\end{flushleft}
\label{t-2}
 \end{table}
\begin{table}[!ht]
\caption{
{\bf Cone Parameters for LMS Images}}
\footnotesize
\centering
\begin{tabular}{|l|l|l|}
\hline
Cone Parameter & Value & Source	\\ \hline\hline
Outer Segment Length & $33\mu m$ & \cite{Rodieck:1998} Appendix B \\ \hline
Inner Segment Length & $2.3\mu m$ &  \cite{Rodieck:1998} Appendix B	 \\ \hline
Specific Density & $0.5$ (axial optical density) & \cite{Rodieck:1998} Appendix B  \\ \hline
Lens Transmittance & See CVRL database & \cite{Stockman:Sharpe:1999}  \\ \hline
Macular Transmittance & See CVRL database & \cite{Bone:Landrum:1992}  \\ \hline
Pupil Diameter & See \cite{Pokorny:Smith:1997b}, Eq.1 & \cite{Pokorny:Smith:1997b} \\ \hline
Eye Length & $16.1mm$ &  \cite{Rodieck:1998} Appendix B	 \\ \hline
Photoreceptor Nomogram & See CVRL database & \cite{Stockman:Sharpe:2000} \\ \hline
Photoreceptor Quantal Efficiency & 0.667 & \cite{Rodieck:1998} page 472 \\ \hline
\end{tabular}
\begin{flushleft} CVRL database is accessible at \emph{http://www-cvrl.ucsd.edu/}.
\end{flushleft}
\label{t-3}
 \end{table}

\bibliography{idb_plos_one}

\end{document}